\documentclass[]{aastex7}

\shorttitle{Tides Tighten the Hycean Habitable Zone}
\shortauthors{Livesey, Becker \& Widicus Weaver}

\graphicspath{{./}{figures/}}

\usepackage[italicdiff]{physics}
\usepackage{mhchem}

\def\msun{{M_\odot}}
\def\rsun{{R_\odot}}
\def\mearth{{M_\oplus}}
\def\rearth{{R_\oplus}}
\def\teff{{T_\text{eff}}}
\def\teq{{T_\text{eq}}}

\def\dist{{\mathcal{R}}}
\newcommand{\rms}[1]{{#1_\mathrm{rms}}}
\newcommand{\revone}[1]{#1}
\newcommand{\revtwo}[1]{#1}

\DeclareMathOperator{\sgn}{sgn}

\begin{document}

\title{Tides Tighten the Hycean Habitable Zone}

\correspondingauthor{Joseph R. Livesey}
\email{jrlivesey@wisc.edu}

\author[0000-0003-3888-3753,gname=Joseph,sname=Livesey]{Joseph R. Livesey}
    \affiliation{Department of Astronomy, University of Wisconsin--Madison, 475 N Charter St, Madison, WI 53706, USA}
    \affiliation{Wisconsin Center for Origins Research, University of Wisconsin--Madison, 475 N Charter St, Madison, WI 53706, USA}
    \email{jrlivesey@wisc.edu}
\author[0000-0002-7733-4522,gname=Juliette,sname=Becker]{Juliette Becker}
    \affiliation{Department of Astronomy, University of Wisconsin--Madison, 475 N Charter St, Madison, WI 53706, USA}
    \affiliation{Wisconsin Center for Origins Research, University of Wisconsin--Madison, 475 N Charter St, Madison, WI 53706, USA}
    \email{juliette.becker@wisc.edu}
\author[0000-0001-6015-3429,gname=Susanna,sname=Widicus Weaver]{Susanna L. Widicus Weaver}
    \affiliation{Department of Astronomy, University of Wisconsin--Madison, 475 N Charter St, Madison, WI 53706, USA}
    \affiliation{Wisconsin Center for Origins Research, University of Wisconsin--Madison, 475 N Charter St, Madison, WI 53706, USA}
    \affiliation{Department of Chemistry, University of Wisconsin--Madison, 1101 University Ave, Madison, WI 53706, USA}
    \email{slww@wisc.edu}

\begin{abstract}
Hycean planets --- exoplanets with \revtwo{substantial water ice layers, deep surface oceans, and hydrogen-rich atmospheres} --- are thought to be favorable environments for life. Due to a relative paucity of atmospheric greenhouse gases, hycean planets have been thought to have wider habitable zones than Earth-like planets, extending down to a few times $10^{-3}$ au for those orbiting M dwarfs. In this Letter, we reconsider the hycean habitable zone accounting for star--planet tidal interaction. We show that for a moderately eccentric hycean planet, the surface temperature contribution from tidal heating truncates the habitable zone at significantly larger orbital radii, and that moderate eccentricity is readily obtained from any massive outer companion in the system. \revone{Though few current hycean planet candidates orbit stars of low enough mass for tides to plausibly significantly alter the extent of the habitable zone, this effect will be important to note as more such candidates are identified orbiting M dwarfs.} We suggest that tides are a significant factor both for determining the extent of the hycean habitable zone around low-mass stars and for the development of a detectable hycean biosphere.
\end{abstract}


\section{Introduction}
\label{sec:intro}

Many studies have parameterized the habitable zone (HZ) for terrestrial exoplanets \citep[e.g.,][]{Kasting1993, Lammer2009, Kopparapu2013}. The exact HZ boundaries can vary based on key characteristics such as stellar host type \citep{Zhan2024}, planetary mass \citep{Kopparapu2014}, atmospheric composition \citep{Kuzucan2025}, and more \citep{Innes2023}. 
These efforts are often driven by the goal of identifying whether newly discovered exoplanets might be considered habitable \citep{Kane2016} and predicting future fruitful observational directions for finding habitable planets \citep{Kane2024}. 

\revone{One interesting recent target for habitability studies have been sub-Neptunes (planets with mass smaller than Neptune) with substantial water layers and hydrogen-dominated atmospheres. These planets were termed ``hycean planets'' by \citet{Madhusudhan2021}.}
Using simple interior models of hycean planets and their atmospheres, \citet{Madhusudhan2021} found that habitable environments can potentially exist on these worlds for a much broader range of equilibrium temperatures than for ``Earth-like'' planets. 
Compared to the Earth-like habitable zone, the hycean habitable zone (HHZ) reaches both smaller semi-major axes and potentially extends to include even unbound planetary orbits. Closer orbital proximity is enabled by the near-absence of greenhouse gases other than \ce{H2} (with water vapor-driven runaway greenhouse occurring only on very hot planets; \citealt{Pierrehumbert2023}) and high albedo. Internal heating from radiogenic sources combined with high pressures and a resultant liquid water layer due to large planet mass allows the HHZ to extend outward indefinitely.

Using these constraints, \citet{Madhusudhan2021} find that planetary equilibrium temperatures below 430 K can yield habitable surfaces for hycean worlds. 
In addition, they identify a ``dark hycean HZ'' interior to this limit, in which a \revone{tidally locked} hycean planet could host a habitable environment on its nightside, assuming \revone{in}efficient redistribution of energy \revone{from dayside to nightside}. The dark HHZ extends up to an equilibrium temperature of 510 K. 
Notably, the inner limit of the HHZ and the dark HHZ extends down to semi-major axes of $\sim 0.009$ au and $\sim 0.006$ au, respectively.

Unlike Earth, short-period exoplanets experience significant tidal interactions with their host stars \citep{Seligman2024}. The heat resulting from tidal dissipation in these planets' interiors can render a significant contribution to the planetary surface temperature, potentially limiting the inner edge of the HZ and extending its outer edge \citep[e.g.,][]{Barnes2013, BarnesHeller2013, Barnes2017}. This effect can be important for planets orbiting M dwarfs, for which the HZ is small and overlaps the regime of significant tides. It is more pronounced in cases where an outer companion buoys the orbital eccentricity and prolongs tidal evolution \citep[e.g.,][]{Livesey2024, Barnes2025}.

Adding to the peril of tidal heating for hycean worlds, these planets can have very high tidal dissipation functions $Q^{-1}$. The classically accepted quality factor for a terrestrial planet is $Q = 100$, and due to efficient dissipation in its ocean Earth has $Q = 12$. Planets that comprise mostly ice, however, can achieve quality factors down to $Q \simeq 2$, depending on the tidal frequency \citep{Tobie2019}. Hycean planets, harboring thick ice layers with $\sim 10^{1-3}$ km overlying oceans \citep{RigbyMadhusudhan2024}, may experience even more rapid dissipation, and thus very high levels of heating from tidal dissipation. 


In this Letter, we examine how tidal heating can influence the location and nature of the habitable zone for hycean planets. In Section \ref{sec:tides}, we calculate the tidal heat generated in a hypothetical hycean planet near the inner edge of the HHZ under the equilibrium tide. We explore in Section \ref{sec:secular} the circumstances under which a companion planet can drive the hycean planet to eccentricities high enough to sustain this heating. In Section \ref{sec:forced-heat} we derive the total increase in equilibrium temperature resulting from tides. Finally, we conclude in Section \ref{sec:discussion} by considering the implications of these results.

\section{Tidal Heating at the Inner Edge of The Habitable Zone} \label{sec:tides}
To compute the heat generated in the planetary interior as a result of tidal forcing, we adopt the constant-phase-lag (CPL) formulation of the equilibrium tide \citep{GoldreichSoter1966, Ferraz-Mello2008}, and assume that the planet is a synchronous rotator with zero obliquity.\footnote{Note that the CPL model is only of second order in the orbital eccentricity of the planet. Therefore, the model \revone{is useful only for} orbital eccentricities $e \lesssim 0.5$, beyond which point the CPL tidal power becomes a significant underestimate \citep[see e.g.,][]{Barnes2020}.} The energy generated by tidal dissipation is then
\begin{align}
    \dot{E}_p &= -\tfrac{1}{8} Z_p \left [ \tfrac{49}{2} \sgn(2\omega_p - 3n) - \tfrac{1}{2} \sgn(2\omega_p - n) - 3 \right ] e^2, \label{eq:tide-power}
\end{align}
where $\omega_p$ is the angular rotational frequency, $n = \sqrt{GM / a^{3}}$ the mean motion, and the strength of the tidal response is parameterized by
\begin{equation}
    Z_p \equiv 3 (G m_\star)^{3/2} (m_p + m_\star) \frac{R_p^5}{a^{15/2}} \frac{k_{2,p}}{Q_p},
\end{equation}
with $R_p$ denoting the planet's physical radius, $a$ its semi-major axis, $k_{2,p}$ its tidal Love number, and $Q$ its tidal quality factor. Equations for $\dot{a}$, $\dot{e}$, and $\dot{\omega}_p$ also come from the CPL model and must be simultaneously solved to yield the full equations of motion, but we concern ourselves here only with the instantaneous value of $\dot{E_p}$.
Equation (\ref{eq:tide-power}) combines the internal energy gained from (i) the loss of orbital kinetic energy and (ii) the loss of rotational kinetic energy. The energy received via instellation is 
\begin{align} \label{eq:instellation}
    \dot{E}_\text{rad} = \frac{\pi (1 - A)}{\sqrt{1 - e^2}} \left ( \frac{R_\star R_p}{a} \right )^2 \sigma \teff^4
\end{align}
where $A$ is the planet's Bond albedo \citep[which we set to 0.5 for a fiducial hycean planet, following][]{Madhusudhan2021}, and the energy is increased by a factor of $(1 - e^2)^{-1/2}$ compared to a planet on a circular orbit \citep{Adams2006, Gallo2024}. 
The equilibrium temperature of the planet is obtained by setting equal the total incoming energy (instellation + tides) and the outgoing longwave radiation:
\begin{align} \label{eq:energy-balance}
    4\pi R_p^2 \sigma \teq^4 = \dot{E}_\text{rad} + \dot{E}_\text{tide},
\end{align}
with the power $\dot{E}_\text{rad}$ given by Equation (\ref{eq:instellation}) and $\dot{E}_\text{tide}$ given by Equation (\ref{eq:tide-power}). 

\begin{figure}
    \centering
    \includegraphics[width=\linewidth]{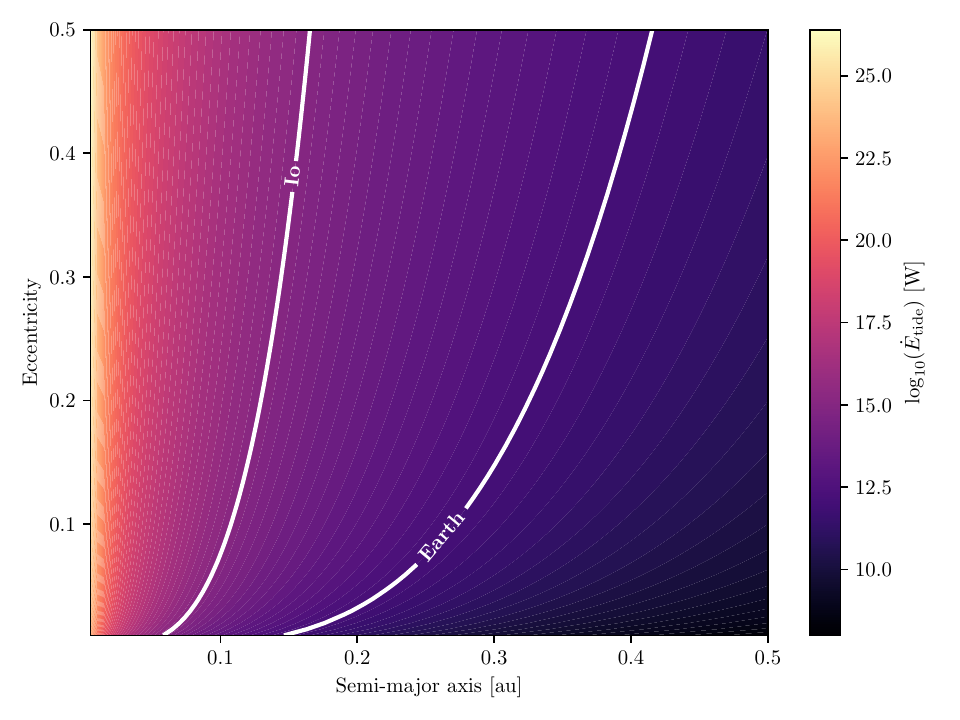}
    \caption{$\dot{E}_\text{tide}$ as a function of semi-major axis and eccentricity for a hycean, synchronously rotating planet with $k_2 = 0.3$, $Q = 2$, $m_p = 7\mearth$, and $R_p = 1.7\rearth$ orbiting a $0.12\msun$ star. The contours correspond to the total tidal heating of Io due to Jupiter ($\sim 10^{15}$ W) and that of Earth due to the Moon ($\sim 10^{12}$ W).}
    \label{fig:tide-power}
\end{figure}

\begin{figure}
    \centering
    \includegraphics[width=\linewidth]{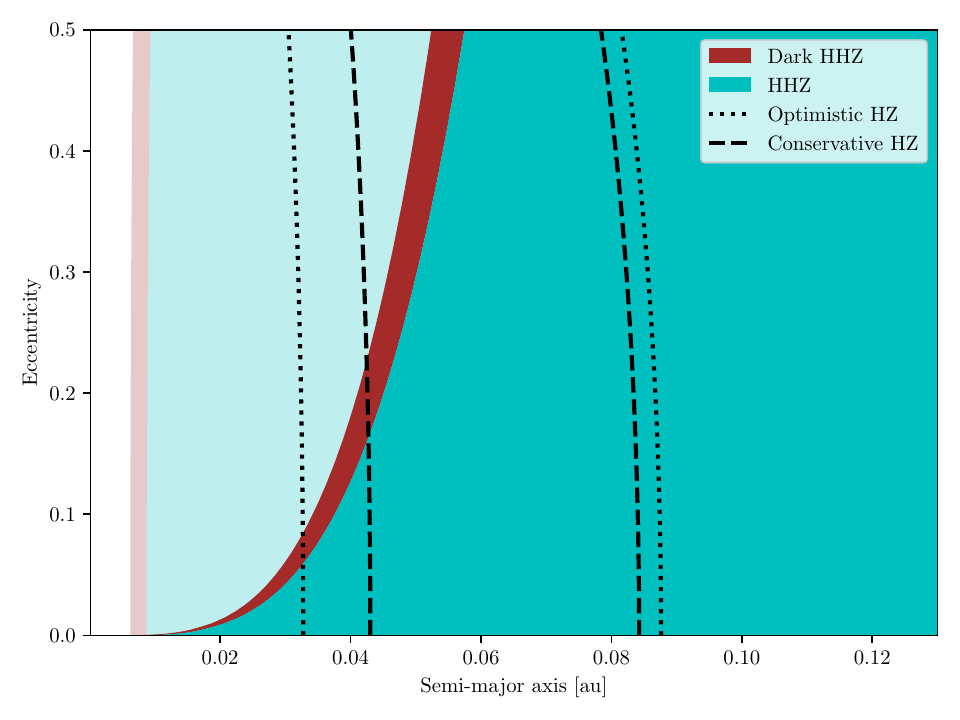}
    \caption{The HHZ (blue shaded regions) and dark HHZ (red shaded regions) around a $0.12\msun$ star for a $7\mearth$, $1.7\rearth$ hycean planet with tidal quality factor $Q = 2$. The low-opacity contours show the habitable zone locations without tidal heating \citep[cf.][]{Madhusudhan2021}, and the high-opacity contours show the habitable zone location for our model where tidal heating is included. The dotted and dashed lines denote the optimistic and conservative HZ limits respectively for Earth-like planets according to the model of \citet{Kopparapu2013}. The conservative HZ is bounded by their ``Recent Venus'' and ``Early Mars'' cases, while the optimistic HZ is bounded by their ``Runaway Greenhouse'' and ``Maximum Greenhouse'' cases.}
    \label{fig:hz-limits}
\end{figure}

In Figure \ref{fig:tide-power}, we plot $\dot{E}_\text{tide}(a, e)$ for a nominal hycean planet, with a low tidal quality factor of $Q = 2$. For this calculation, we adopt a stellar mass of $0.12\msun$, a stellar radius of $0.15\rsun$, a stellar effective temperature of 2992 K, a planetary mass of $7\mearth$, and a planetary radius of $1.7\rearth$. These values roughly correspond to Proxima Centauri --- a representative late M dwarf --- and the candidate hycean exoplanet LHS 1140 b.
In Figure \ref{fig:hz-limits} we plot limits of the HHZ accounting for tides on the same planet. The inner limits computed by \citet{Madhusudhan2021} are at equilibrium temperatures of 430 K for \revone{hycean planets that are either nonsynchronously rotating or are tidally locked but have have efficient redistribution of heat from the planetary dayside to nightside} and 510 K for those that are tidally locked \revone{and inefficiently redistribute heat} (dark hyceans, \revone{where the nightside could be habitable}). Marked in light (bold) colors are the HHZ regimes without (with) tides. The overlying curves are the limits of the optimistic and conservative HZs for Earth-like planets, as computed by \citet{Kopparapu2013}. We see that at moderate eccentricities the inner limit of the HHZ traverses the inner HZ limits for ``Earth-like'' planets.

\begin{figure}
    \centering
    \includegraphics[width=\linewidth]{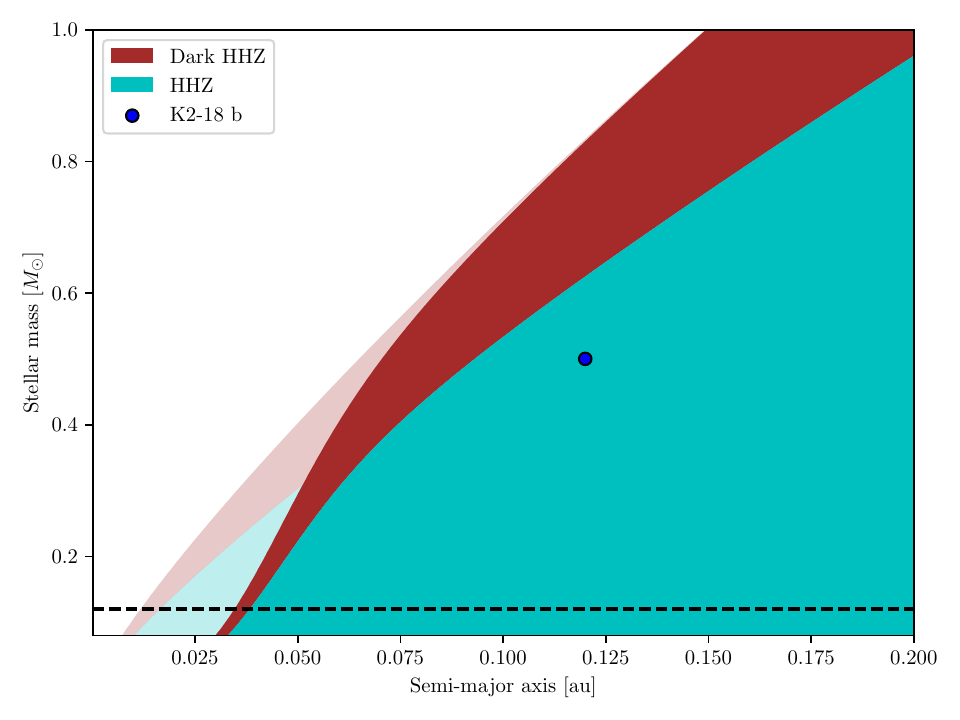}
    \caption{\revone{Similarly to Figure \ref{fig:hz-limits}, the HHZ and dark HHZ around stars of various mass for a $7\mearth$, 1.7$\rearth$ hycean planet with tidal quality factor $Q = 2$. Here we use an orbital eccentricity $e = 0.1$ throughout: about the value for the hycean candidate planet K2-18 b, whose location in this parameter space is indicated by the blue dot. The dashed line at $0.12\msun$ points out the stellar mass used in Section \ref{sec:tides} and Figure \ref{fig:hz-limits}. Clearly, the effect of tides on the extent of the habitable zone becomes negligible at high stellar masses.}}
    \label{fig:hz-lim-stellarmass}
\end{figure}

\revone{We emphasize that our analysis here focuses on hycean planet orbiting very low-mass stars. This is not the situation of most current hycean candidates. In Figure \ref{fig:hz-lim-stellarmass} we show these habitable zones for a hycean planet, also with $Q = 2$, orbiting stars of various mass between 0.08 and 1 $\msun$.}

\section{Forced Eccentricity From a Companion} \label{sec:secular}
While a lone short-period planet will quickly circularize its orbit through tides, the presence of a large outer companion can sustain a significant eccentricity over long timescales, substantially prolonging the duration of tidal heating. A companion that is in mean motion resonance (MMR) with the planet in question will resonantly force its eccentricity, achieving such an effect \citep[e.g.,][]{Tamayo2017}. However, even in the absence of MMRs, a companion can boost the eccentricity of the inner planet through secular interactions. \revone{Here we consider an outer companion that is much more massive than the inner hycean planet. These outer companions do occur in planetary systems around M dwarfs; the occurrence rate of giants in such systems has been found to be $\sim 10$\% \citep{Bryant2023}, and the occurrence rate of planets in the range 10--100$\mearth$ is $\sim 20$\% \citep{Sabotta2021}.}

From Equation (\ref{eq:tide-power}), we see that for energy to be dissipated within a planetary interior via the equilibrium tide, the planetary orbit must have a non-zero orbital eccentricity. 
While tidal forces generally operate to circularize the orbits of short-period planets, non-zero planetary eccentricities can be maintained in tightly packed multi-planet systems through forced eccentricity induced by planet--planet interactions \citep{Peale1979, Barr2018, Peterson2023}. 

Assume a hycean planet experiencing significant tides is perturbed by an outer companion. In the absence of any MMR, the disturbing function describing this perturbation to second order in $e$ is
\begin{equation} \label{eq:disturbing}
    \dist = \frac{G m'}{8a} \alpha^2 \left [ b^{(1)}_{3/2}(\alpha) e^2 - 2b^{(2)}_{3/2}(\alpha) e e' \cos(\varpi - \varpi') \right ],
\end{equation}
where primed quantities relate to the outer planet, $\alpha = a/a'$, and the Laplace coefficients are given by
\begin{equation} \label{eq:laplace-coefficient}
    b^{(\ell)}_s(\alpha) = \frac{2}{\pi} \int_0^\pi \frac{\cos(\ell \psi) \: d\psi}{[1 + \alpha^2 - 2\alpha \cos(\psi)]^s}.
\end{equation}
For this calculation, we have assumed that the outer planet is sufficiently large relative to the inner planet that its orbit is time-independent ($\dist' = \text{constant}$). We adopt a reference frame co-rotating with the outer planet's periapse and set $\varpi' = 0$. Substituting $h = e \sin(\varpi)$ and $k = e \cos(\varpi)$ yields
\begin{equation}
    \dist = \frac{G m'}{8a} \alpha^2 b^{(1)}_{3/2}(\alpha) \left [ h^2 + k^2 - 2e' \frac{b^{(2)}_{3/2}(\alpha)}{b^{(1)}_{3/2}(\alpha)} k \right ].
\end{equation}
Hamilton's equations 
for this system are $\dot{h} = (n a^2)^{-1} \pdv*{\dist}{k}$ and $\dot{k} = -(n a^2)^{-1} \pdv*{\dist}{h}$, where $n = \sqrt{G m_\star / a^3}$ is the mean motion \citep{MurrayDermott1999}. We obtain
\begin{align}
    \dot{h} &= -\xi [k + \Phi] \\
    \dot{k} &= -\xi h,
\end{align}
where we have defined $\xi \equiv \frac{1}{4} G^{1/2} m' m_\star^{-1/2} a^{-3/2} \alpha^2 b^{(1)}_{3/2}(\alpha)$ and $\Phi \equiv -e' b^{(2)}_{3/2}(\alpha) / b^{(1)}_{3/2}(\alpha)$. Solving these coupled differential equations shows that the eccentricity evolves as a driven harmonic oscillator.
\begin{align}
    h(t) &= h_0 \cos(\xi t) \label{eq:h} \\
    k(t) &= (k_0 + \Phi) \cos(\xi t) - \Phi \label{eq:k}
\end{align}
Using Equations (\ref{eq:h}) and (\ref{eq:k}), we can compute the average of $e^2 = h^2 + k^2$ over a full secular forcing period, $2\pi/\xi$, and thus the root-mean-square value of the eccentricity $\rms{e}$.
\begin{align}
    \rms{e} &= \left ( \frac{\xi}{2\pi} \int_0^{2\pi/\xi} (h^2 + k^2) \: dt \right )^{1/2} = \left ( \tfrac{1}{2} [h_0^2 + (k_0 + \Phi)^2 + 2\Phi^2] \right )^{1/2}
\end{align}

\begin{figure}
    \centering
    \includegraphics[width=\linewidth]{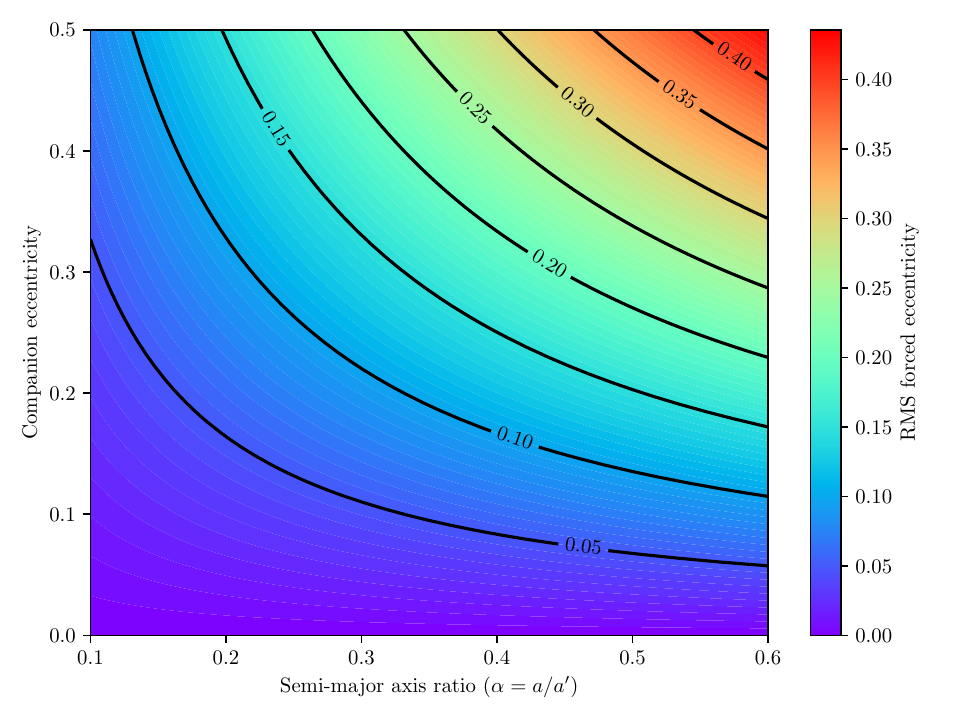}
    \caption{The root-mean-square forced eccentricity $\rms{e}$ for a planet with zero free eccentricity and accompanied by an outer companion in the secular formulation. \revone{Note that $\rms{e}$ does not depend upon the mass of the inner planet in this simple model.}}
    \label{fig:forced-rms-e}
\end{figure}

This relationship gives us $\rms{e}$ given the companion's orbit (via $\Phi$) and the planet's free eccentricity ($h_0^2 + k_0^2$), which we take to be set by the tidal evolution described in Section \ref{sec:tides}. Note that even once the free eccentricity has been damped to zero by tides $\rms{e} = \sqrt{3/2} \Phi$, which is always finite.\footnote{Note that $\rms{e}$ does not depend upon the mass of the companion, nor upon the semi-major axes of the planets (only their ratio $\alpha$). These quantities influence the secular frequency $\xi$, but averaging over all time removes the frequency dependence.}
Hence tidal heating can persist indefinitely in the presence of a massive companion. The relationship among $\rms{e}$, $e'$, and $\alpha$ is visualized in Figure \ref{fig:forced-rms-e}.

The value of $\rms{e}$ sets the average tidal heat on a given planet whose orbit is circularized by tides, and thus the limits on the HHZ for that world, as given in Figures \ref{fig:tide-power} and \ref{fig:hz-limits}. From Figure \ref{fig:hz-limits}, we see that even for low sustained eccentricities, $e \gtrsim 10^{-3}$, the inner edge of the HHZ is pushed significantly outward. Figure \ref{fig:forced-rms-e} shows that on average, we expect a short-period planet with a companion to surpass this threshold even for fairly low $e'$ and $\alpha$; we should therefore in general expect significant and sustained tidal heating of these worlds.

\section{Tidal Heating From the Forced Eccentricity} \label{sec:forced-heat}

The simplicity of both the CPL model of the equilibrium tide detailed in Section \ref{sec:tides} and the two-planet secular system developed in Section \ref{sec:secular} allow us to write down a closed-form expression for the factor by which the planetary equilibrium temperature increases from its instellation-only value when we account for the contribution of tidal power. Consider a synchronously rotating planet with zero obliquity and zero free eccentricity. Equation (\ref{eq:tide-power}) for this planet becomes the following.
\begin{align}
    \dot{E}_\text{tide} = \tfrac{7}{2} Z_p e^2
\end{align}
If we assume that $\dot{a}$ is negligible, then $Z_p$ is a constant. The average tidal power provided to the planet over a secular cycle is
\begin{align}
    \langle \dot{E}_\text{tide} \rangle = \tfrac{21}{4} Z_p \Phi^2.
\end{align}
Meanwhile, the heat due to instellation alone (Equation \ref{eq:instellation}) is
\begin{align}
    \dot{E}_\text{rad} \simeq \pi (1 - A) \left ( \frac{R_\star R_p}{a} \right )^2 \sigma \teff^4 \left ( 1 + \tfrac{1}{2} e^2 \right ),
\end{align}
which averaged in the same way yields
\begin{align}
    \langle \dot{E}_\text{rad} \rangle \simeq \pi (1 - A) \left ( \frac{R_\star R_p}{a} \right )^2 \sigma \teff^4 \left ( 1 + \tfrac{3}{4} \Phi^2 \right ).
\end{align}
We can now find the overall factor by which the planetary equilibrium temperature increases when we account for tides:
\begin{align} \label{eq:temp-ratio-e}
    \left ( \frac{T_\text{eq}^\text{rad+tide}}{T_\text{eq}^\text{rad}} \right )^4 = \frac{\dot{E}_\text{rad} + \dot{E}_\text{tide}}{\dot{E}_\text{rad}} \simeq 1 + \frac{7Z_p}{\pi (1 - A) \sigma \teff^4} \left ( \frac{a}{R_\star R_p} \right )^2 \frac{e^2}{2 + e^2}.
\end{align}
Not surprisingly, the physical and orbital properties of the inner planet and the star set the magnitude of the increase in equilibrium temperature when we account for the effect of tides. This formula yields in general the amount by which the equilibrium temperature is underestimated without considering tidal heating (assuming the heating is well described by the equilibrium tide).

We can likewise determine the factor by which the planetary equilibrium temperature increases --- averaged over all time --- when we account for tides. 
\begin{align} \label{eq:temp-ratio-phi}
    \frac{\langle (T_\text{eq}^\text{rad+tide})^4 \rangle}{\langle (T_\text{eq}^\text{rad})^4 \rangle} = \frac{\langle \dot{E}_\text{rad} \rangle + \langle \dot{E}_\text{tide} \rangle}{\langle \dot{E}_\text{rad} \rangle} \simeq 1 + \frac{21Z_p}{\pi (1 - A) \sigma \teff^4} \left ( \frac{a}{R_\star R_p} \right )^2 \frac{\Phi^2}{4 + 3\Phi^2}
\end{align}
For a planet with zero free eccentricity experiencing secular interactions with a massive outer companion, this formula gives the \revone{ratio between secular cycle-averaged temperatures.} 
For ratios close to unity tidal effects can be neglected; for larger ratios, to fully model the state of the planet tidal effects must be included.  

Importantly, Equations (\ref{eq:temp-ratio-e}) and (\ref{eq:temp-ratio-phi}) give a quantitative determination of the effect of tides on equilibrium temperature for a planet of any given mass and radius, orbiting a star of any given mass (via $Z_p$). 


\section{Discussion} \label{sec:discussion}
In this work we have considered the effects of tides on hycean planets orbiting small stars. These worlds are excellent targets for the search for biosignatures in transmission spectroscopy, due to their extended atmospheres and potential for habitability \citep{PierrehumbertGaidos2011, Madhusudhan2020}. In particular, \citet{Madhusudhan2021} show that the HZ of hycean worlds (the HHZ) far exceeds that of Earth-like planets with atmospheres of higher mean molecular weight; down to semi-major axes of $\sim 10^{-3}$ au and out indefinitely far from the host star. We have shown that when the effect of tides is considered the inner limits of the HHZ are not necessarily as optimistic as those calculated by \citet{Madhusudhan2021} for small host stars. Increasing the stellar mass, however, quickly reduces this effect. G, K, and early M dwarf host stars will have HHZs at larger semi-major axes, which means planets in those zones will be largely unaffected by tides.

Hycean planets are likely to exhibit stronger tidal responses than a fiducial terrestrial world; at moderate eccentricities, the HHZ actually begins at higher semi-major axes than the Earth-like HZ. We expect tides to have little effect on a lone planet at such small orbital radii. However, the presence of a large outer companion with moderate eccentricity will force an eccentricity cycle that periodically and indefinitely heats the interior of the planet in question (Figure \ref{fig:tide-power}), and push out the inner boundaries of the HHZ (Figure \ref{fig:hz-limits}). In the secular picture of planet--planet interactions, the magnitude of this eccentricity cycle is set by the planet--planet separation and the orbit of the outer planet (Figure \ref{fig:forced-rms-e}). 
Orbital resonance is another mechanism that can sustain moderate to high eccentricities over long timescales \citep[e.g.,][]{MurrayDermott1999, Ketchum2011, MacDonald2023}. In some cases, the regular conjunctions provided by mean motion resonance can lead to larger allowable eccentricities than standard secular dynamical stability criteria \citep{Tamayo2017, Petit2020, Lammers2024}. 
Because of this, in systems containing hycean planets in mean motion resonance with a nearby companion, larger eccentricities than those shown in Figure \ref{fig:forced-rms-e} may be possible and stable over long timescales.

In this work, we show that the dark HHZ’s inner boundary expands to larger orbital radii for moderate orbital eccentricities (up to $e=0.5$). In addition to this limitation, previous studies \citep{Selsis2013, Bolmont2016} indicate that for $e > 0.7$ a tidally locked planet’s nightside will lack continuous darkness, making the dark HHZ undefined. 
An interesting direction for future research would be to determine whether tidal heating might elevate planetary temperatures enough for hycean planets to enter a tidal greenhouse state \citep{Pierrehumbert2023}, inflating their radii and altering their tidal response.

We conclude with a note on the significance of this additional heat source for life. A recent possible detection of dimethyl sulfide (DMS) in the atmosphere of the potential hycean exoplanet K2-18 b \citep{Madhusudhan2023} may indicate the presence of ocean-faring life; the only major source of DMS on Earth is phytoplankton. The surface equilibrium temperature of a planet bearing simple unicellular life has a direct effect on the evolutionary rate --- and hence the biological complexity --- of that life \citep{Petraccone2024}. In particular, a hycean planet with a surface equilibrium temperature 10 K higher than the Earth median, and hosting the same last universal common ancestor, could see key unicellular clades (i.e., DMS-producing organisms) arise $\sim 2.5$ Gyr earlier than on Earth, based on a simple parameterization of evolutionary rates \citep{MitchellMadhusudhan2025}. On bodies with deep oceans, tides will deliver power directly to the ocean \citep[on Earth most tidal energy is dissipated in the ocean, primarily through bottom drag in shallow water and through the scattering of surface tides into internal waves in deep water;][]{MunkMacdonald1960, EgbertRay2000}. We suggest, therefore, that strong tides on hycean planets could yield a significant power source for life and ultimately accelerate biological evolution.

\begin{acknowledgments}
The authors thank Abygail Waggoner and Hannah Zanowski for useful conversations. \revone{We are grateful to the anonymous reviewer for their suggestions, which improved the clarity of this manuscript.}

Resources and financial support for this project were provided in part by the Wisconsin Center for Origins Research (WiCOR) at the University of Wisconsin--Madison.
This work was supported by the University of Wisconsin--Madison Research Forward program sponsored by the Office of the Vice Chancellor for Research (OVCR) through funding provided by the Wisconsin Alumni Research Foundation (WARF).

\end{acknowledgments}

\software{Astropy \citep{astropy2022}, Matplotlib \citep{Hunter2007}, Numpy \citep{Harris2020}}

\bibliography{references}{}
\bibliographystyle{aasjournal}

\end{document}